\batchmode
\makeatletter
\def\input@path{{\string"C:/Users/anuppari/Copy/Lab/NCR/[TAC] Integral CL/\string"/}}
\makeatother
\documentclass{IEEEtran}
\pdfoutput=1
\usepackage[T1]{fontenc}
\usepackage[latin9]{inputenc}
\usepackage{verbatim}
\usepackage{amsmath}
\usepackage{amsthm}
\usepackage{amssymb}
\usepackage{stackrel}
\usepackage{graphicx}
\usepackage{esint}

\makeatletter

\providecommand{\tabularnewline}{\\}

\theoremstyle{definition}
\newtheorem{assumption}{Assumption}
  \theoremstyle{plain}
  \newtheorem{thm}{\protect\theoremname}

\usepackage{cite}
\IEEEoverridecommandlockouts

\makeatother

\providecommand{\theoremname}{Theorem}

\begin{document}

\title{Integral Concurrent Learning: Adaptive Control with Parameter Convergence
without PE or State Derivatives\thanks{$^{1}$Department of Mechanical and Aerospace Engineering, University
of Florida, Gainesville FL 32611-6250, USA Email:\{anuppari, rkamalapurkar,
wdixon\}@ufl.edu}\thanks{This research is supported in part by a Task Order contract with the
Air Force Research Laboratory, Munitions Directorate at Eglin AFB,
and Office of Naval Research Grant N00014-13-1-0151. Any opinions,
findings and conclusions or recommendations expressed in this material
are those of the author(s) and do not necessarily reflect the views
of the sponsoring agency.}}

\author{Anup Parikh$^{1}$, Rushikesh Kamalapurkar$^{1}$, Warren E. Dixon$^{1}$}
\maketitle
\begin{abstract}
Concurrent learning is a recently developed adaptive update scheme
that can be used to guarantee parameter convergence without requiring
persistent excitation. However, this technique requires knowledge
of state derivatives, which are usually not directly sensed and therefore
must be estimated. A novel integral concurrent learning method is
developed in this paper that removes the need to estimate state derivatives
while maintaining parameter convergence properties. A Monte Carlo
simulation illustrates improved robustness to noise compared to the
traditional derivative formulation.\end{abstract}

\begin{IEEEkeywords}
Adaptive Control, Parameter Estimation
\end{IEEEkeywords}

\section{Introduction\label{sec:Introduction}}

Adaptive control methods provide a means to achieve a control objective
despite uncertainties in the system model. Adaptive estimates are
developed through insights from a Lyapunov-based analysis as a means
to yield a desired objective. Although a regulation or tracking objective
can be achieved with this scheme, it is well known that the parameter
estimates may not approach the true parameters using a least-squares
or a gradient based online update law without persistent excitation
\cite{Ioannou1996,Narendra1989,Sastry1989a}. However, the persistence
of excitation condition cannot be guaranteed \textit{a priori} for
nonlinear systems, and is difficult to check online, in general.

Motivated by the desire to learn the true parameters, or at least
to gain the increased robustness and improved transient performance
that parameter convergence provides (see \cite{Duarte.Narendra1989,Krstic.Kokotovic.ea1993,Chowdhary.Johnson2011a}),
a new adaptive update scheme known as concurrent learning (CL) was
recently developed in the pioneering work of \cite{Chowdhary.Johnson2011a,Chowdhary2010a,Chowdhary.Yucelen.ea2012}.
The principle idea of CL is to use recorded input and output data
of the system dynamics to apply batch-like updates to the parameter
estimate dynamics. These updates yield a negative definite, parameter
estimation error term in the stability analysis, which allows parameter
convergence to be established provided a finite excitation condition
is satisfied. The finite excitation condition is a weaker condition
than persistent excitation (since excitation is only needed for a
finite amount of time) and can be checked online by verifying the
positivity of the minimum singular value of a function of the regressor
matrix. However, all current CL methods require that the output data
include the state derivatives, which may not be available for all
systems. Since the naive approach of finite difference of the state
measurements leads to noise amplification, and since only past recorded
data, opposed to real-time data, is needed for CL, techniques such
as online state derivative estimation or smoothing have been employed,
e.g., \cite{Muehlegg.Chowdhary.ea2012,arXivKamalapurkar.Reish.easubmitted}.
However, these methods typically require tuning parameters such as
an observer gain, switching threshold, etc. in the case of the online
derivative estimator, and basis, basis order, covariance, time window,
etc. in the case of smoothing, to produce satisfactory results.

In this note, we reformulate the CL method in terms of an integral,
removing the need to estimate state derivatives. Also, the only additional
tuning parameter beyond what is needed for gradient-based adaptive
control designs is the time window of integration, which is analogous
to the smoothing buffer window that is already required for smoothing
based techniques. Despite the reformulation, the stability results
still hold (i.e., parameter convergence) and Monte Carlo simulation
results suggest greater robustness to noise compared to derivative
based CL implementations.

\section{Control Objective\label{sec:Control-Objective}}

To illustrate the integral CL method, consider an example dynamic
system modeled as
\begin{equation}
\dot{x}\left(t\right)=f\left(x\left(t\right),t\right)+u\left(t\right)\label{eq:dynamics}
\end{equation}
where $t\in[0,\infty)$, $x:[0,\infty)\rightarrow\mathbb{R}^{n}$
are the measureable states, $u:[0,\infty)\rightarrow\mathbb{R}^{n}$
is the control input and $f:\mathbb{R}^{n}\times[0,\infty)\rightarrow\mathbb{R}^{n}$
represents the locally Lipschitz drift dynamics, with some unknown
parameters. In the following development, as is typical in adaptive
control, $f$ is assumed to be linearly parametrized in the unknown
parameters, i.e., 
\begin{equation}
f\left(x,t\right)=Y\left(x,t\right)\theta\label{eq:linear_parameterization}
\end{equation}
where $Y:\mathbb{R}^{n}\times[0,\infty)\rightarrow\mathbb{R}^{n\times m}$
is a regressor matrix and $\theta\in\mathbb{R}^{m}$ represents the
constant, unknown system parameters. To quantify the state tracking
and parameter estimation objective of the adaptive control problem,
the tracking error and parameter estimate error are defined as
\begin{equation}
e\left(t\right)\triangleq x\left(t\right)-x_{d}\left(t\right)\label{eq:error_definition}
\end{equation}
\begin{equation}
\tilde{\theta}\left(t\right)\triangleq\theta-\hat{\theta}\left(t\right)\label{eq:parameter_estimation_error}
\end{equation}
where $x_{d}:[0,\infty)\rightarrow\mathbb{R}^{n}$ is a known, continuously
differentiable desired trajectory and $\hat{\theta}:[0,\infty)\rightarrow\mathbb{R}^{m}$
is the parameter estimate. In the following, functional arguments
will be omitted for notational brevity, e.g., $x\left(t\right)$ will
be denoted as $x$, unless necessary for clarity.

To achieve the control objective, the following controller is commonly
used:
\begin{equation}
u\left(t\right)\triangleq\dot{x}_{d}-Y\left(x,t\right)\hat{\theta}-Ke\label{eq:control_design}
\end{equation}
where $K\in\mathbb{R}^{n\times n}$ is a positive definite constant
control gain. Taking the time derivative of (\ref{eq:error_definition})
and substituting for (\ref{eq:dynamics}), (\ref{eq:linear_parameterization}),
and (\ref{eq:control_design}), yields the closed loop error dynamics
\begin{eqnarray}
\dot{e} & = & Y\left(x,t\right)\theta+\dot{x}_{d}-Y\left(x,t\right)\hat{\theta}-Ke-\dot{x}_{d}\nonumber \\
 & = & Y\left(x,t\right)\tilde{\theta}-Ke\label{eq:closed_loop_dynamics}
\end{eqnarray}
The parameter estimation error dynamics are determined by taking the
time derivative of (\ref{eq:parameter_estimation_error}), yielding
\begin{equation}
\dot{\tilde{\theta}}\left(t\right)=-\dot{\hat{\theta}}.\label{eq:estimation_error_dynamics}
\end{equation}
 An integral CL-based update law for the parameter estimate is designed
as
\begin{eqnarray}
\dot{\hat{\theta}}\left(t\right) & \triangleq & \Gamma Y\left(x,t\right)^{T}e\label{eq:adaptive_update_law}\\
 &  & +k_{CL}\Gamma\stackrel[i=1]{N}{\sum}\mathcal{Y}_{i}^{T}\left(x\left(t_{i}\right)-x\left(t_{i}-\Delta t\right)-\mathcal{U}_{i}-\mathcal{Y}_{i}\hat{\theta}\right)\nonumber 
\end{eqnarray}
where $k_{CL}\in\mathbb{R}$ and $\Gamma\in\mathbb{R}^{m\times m}$
are constant, positive definite control gains, $N\in\mathbb{Z}^{+}$is
a positive constant, $t_{i}\in\left[0,t\right]$ are time points between
the initial time and the current time, $\mathcal{Y}_{i}\triangleq\mathcal{Y}\left(t_{i}\right)$,
$\mathcal{U}_{i}\triangleq\mathcal{U}\left(t_{i}\right)$,
\begin{equation}
\mathcal{Y}\left(t\right)\triangleq\begin{cases}
0_{n\times m} & t\in\left[0,\:\Delta t\right]\\
\int_{t-\Delta t}^{t}Y\left(x\left(\tau\right),\tau\right)d\tau & t>\Delta t
\end{cases}\label{eq:script_Y_definition}
\end{equation}
\begin{equation}
\mathcal{U}\left(t\right)\triangleq\begin{cases}
0_{n\times1} & t\in\left[0,\:\Delta t\right]\\
\int_{t-\Delta t}^{t}u\left(\tau\right)d\tau & t>\Delta t
\end{cases}\label{eq:script_U_definition}
\end{equation}
$0_{n\times m}$ denotes an $n\times m$ matrix of zeros, and $\Delta t\in\mathbb{R}$
is a positive constant denoting the size of the window of integration.
The concurrent learning term (i.e., the second term) in (\ref{eq:adaptive_update_law})
represents saved data. The principal idea behind this design is to
utilize recorded input-output data generated by the dynamics to further
improve the parameter estimate. See \cite{Chowdhary2010a} for a discussion
on how to choose data points to record.

The integral CL-based adaptive update law in (\ref{eq:adaptive_update_law})
differs from traditional state derivative based CL update laws given
in, e.g., \cite{Chowdhary.Johnson2011a,Chowdhary2010a,Chowdhary.Yucelen.ea2012}.
Specifically, the state derivative, control, and regressor terms,
i.e., $\dot{x}$, $u$, and $Y$, respectively, used in \cite{Chowdhary.Johnson2011a,Chowdhary2010a,Chowdhary.Yucelen.ea2012}
are replaced with the integral of those terms over the time window
$\left[t-\Delta t,\:t\right]$.

Substituting (\ref{eq:linear_parameterization}) into (\ref{eq:dynamics}),
and integrating yields
\[
\int_{t-\Delta t}^{t}\dot{x}\left(\tau\right)d\tau=\int_{t-\Delta t}^{t}Y\left(x,\tau\right)\theta d\tau+\int_{t-\Delta t}^{t}u\left(\tau\right)d\tau,
\]
$\forall t>\Delta t$. Using the Fundamental Theorem of Calculus and
the definitions in (\ref{eq:script_Y_definition}) and (\ref{eq:script_U_definition}),
\begin{equation}
x\left(t\right)-x\left(t-\Delta t\right)=\mathcal{Y}\left(t\right)\theta+\mathcal{U}\left(t\right)\label{eq:integrated_dynamics}
\end{equation}
$\forall t>\Delta t$, where the fact that $\theta$ is a constant
was used to pull it outside the integral. Rearranging (\ref{eq:integrated_dynamics})
and substituting into (\ref{eq:adaptive_update_law}) yields%
\begin{equation}
\dot{\hat{\theta}}\left(t\right)=\Gamma Y\left(x,t\right)^{T}e+k_{CL}\Gamma\stackrel[i=1]{N}{\sum}\mathcal{Y}_{i}^{T}\mathcal{Y}_{i}\tilde{\theta}.\label{eq:simplified_adaptive_update_law}
\end{equation}

\section{Stability Analysis\label{sec:Stability-Analysis}}

To facilitate the following analysis, let $\eta:[0,\infty)\rightarrow\mathbb{R}^{n+m}$
represent a composite vector of the system states and parameter estimation
errors, defined as $\eta\left(t\right)\triangleq\left[\begin{array}{cc}
e^{T} & \tilde{\theta}^{T}\end{array}\right]^{T}$. Also, let $\mbox{\ensuremath{\lambda}}_{\min}\left\{ \cdot\right\} $
and $\mbox{\ensuremath{\lambda}}_{\max}\left\{ \cdot\right\} $ represents
the minimum and maximum eigenvalues of $\left\{ \cdot\right\} $,
respectively.
\begin{assumption}
\label{thm:finite_excitation}The system is sufficiently excited over
a finite duration of time. Specifically, $\exists\underline{\lambda}>0,\:\exists T>\Delta t:\forall t\geq T,\:\mbox{\ensuremath{\lambda}}_{\min}\left\{ \stackrel[i=1]{N}{\sum}\mathcal{Y}_{i}^{T}\mathcal{Y}_{i}\right\} \geq\underline{\lambda}$. \end{assumption}
\begin{thm}
\label{thm:bounded_stability}For the system defined in (\ref{eq:dynamics})
and (\ref{eq:estimation_error_dynamics}), the controller and adaptive
update law defined in (\ref{eq:control_design}) and (\ref{eq:adaptive_update_law})
ensures bounded tracking and parameter estimation errors during the
time interval $t\in\left[0,T\right]$.\end{thm}
\begin{IEEEproof}
Let $V:\mathbb{R}^{n+m}\rightarrow\mathbb{R}$ be a candidate Lyapunov
function defined as
\[
V\left(\eta\right)=\frac{1}{2}e^{T}e+\frac{1}{2}\tilde{\theta}^{T}\Gamma^{-1}\tilde{\theta}.
\]
Taking the derivative of $V$ along the trajectories of (\ref{eq:dynamics})
during $t\in\left[0,T\right]$, substituting the closed loop error
dynamics in (\ref{eq:closed_loop_dynamics}) and the equivalent adaptive
update law in (\ref{eq:simplified_adaptive_update_law}), and simplifying
yields%
\[
\dot{V}\leq-e^{T}Ke,\:\:\forall t\in\left[0,T\right]
\]
which implies the system states remain bounded via \cite[Theorem 4.18]{Khalil2002}.
Further, since $\dot{V}\leq0$, $V\left(\eta\left(T\right)\right)\leq V\left(\eta\left(0\right)\right)$
and therefore $\left\Vert \eta\left(T\right)\right\Vert \leq\sqrt{\frac{\beta_{2}}{\beta_{1}}}\left\Vert \eta\left(0\right)\right\Vert $,
where $\beta_{1}\triangleq\frac{1}{2}\min\left\{ 1,\mbox{\ensuremath{\lambda}}_{\min}\left\{ \Gamma^{-1}\right\} \right\} $
and $\beta_{2}\triangleq\frac{1}{2}\max\left\{ 1,\mbox{\ensuremath{\lambda}}_{\max}\left\{ \Gamma^{-1}\right\} \right\} $
\end{IEEEproof}

\begin{thm}
\label{thm:exponential_stability}For the system defined in (\ref{eq:dynamics})
and (\ref{eq:estimation_error_dynamics}), the controller and adaptive
update law defined in (\ref{eq:control_design}) and (\ref{eq:adaptive_update_law})
ensures globally exponential tracking in the sense that
\begin{equation}
\left\Vert \eta\left(t\right)\right\Vert \leq\left(\frac{\beta_{2}}{\beta_{1}}\right)\exp\left(\lambda_{1}T\right)\left\Vert \eta\left(0\right)\right\Vert \exp\left(-\lambda_{1}t\right),\:\:\forall t\in[0,\infty).\label{eq:eta_trajectory}
\end{equation}
\end{thm}
\begin{IEEEproof}
Let $V:\mathbb{R}^{n+m}\rightarrow\mathbb{R}$ be a candidate Lyapunov
function defined as
\[
V\left(\eta\right)=\frac{1}{2}e^{T}e+\frac{1}{2}\tilde{\theta}^{T}\Gamma^{-1}\tilde{\theta}.
\]
Taking the derivative of $V$ along the trajectories of (\ref{eq:dynamics})
during $t\in[T,\infty)$, substituting the closed loop error dynamics
in (\ref{eq:closed_loop_dynamics}) and the equivalent adaptive update
law in (\ref{eq:simplified_adaptive_update_law}), and simplifying
yields%
\[
\dot{V}=-e^{T}Ke-k_{CL}\tilde{\theta}^{T}\stackrel[i=1]{N}{\sum}\mathcal{Y}_{i}^{T}\mathcal{Y}_{i}\tilde{\theta},\:\:\forall t\in[T,\infty).
\]
From Assumption \ref{thm:finite_excitation}, $\mbox{\ensuremath{\lambda}}_{\min}\left\{ \stackrel[i=1]{N}{\sum}\mathcal{Y}_{i}^{T}\mathcal{Y}_{i}\right\} >0$,
$\forall t\in[T,\infty)$, which implies that $\stackrel[i=1]{N}{\sum}\mathcal{Y}_{i}^{T}\mathcal{Y}_{i}$
is positive definite and therefore $\dot{V}$ is upper bounded by
a negative definite function of $\eta$. Invoking \cite[Theorem 4.10]{Khalil2002},
$e$ and $\tilde{\theta}$ are globally exponentially stable, i.e.,
$\forall t\in[T,\infty)$,
\[
\left\Vert \eta\left(t\right)\right\Vert \leq\sqrt{\frac{\beta_{2}}{\beta_{1}}}\left\Vert \eta\left(T\right)\right\Vert \exp\left(-\lambda_{1}\left(t-T\right)\right)
\]
where $\lambda_{1}\triangleq\frac{1}{\beta_{2}}\min\left\{ \lambda_{min}\left\{ K\right\} ,k_{CL}\underline{\lambda}\right\} $.
The composite state vector can be further upper bounded using the
results of Theorem \ref{thm:bounded_stability}, yielding (\ref{eq:eta_trajectory}).

\end{IEEEproof}

\section{Simulation\label{sec:Simulation}}

A Monte Carlo simulation was performed to demonstrate the application
of the theoretical results presented in Section \ref{sec:Stability-Analysis}
and to illustrate the increased robustness to noise compared to the
traditional state derivative based CL methods. The following example
system was used in the simulations:
\[
\dot{x}\left(t\right)=\left[\begin{array}{cccc}
x_{1}^{2} & \sin\left(x_{2}\right) & 0 & 0\\
0 & x_{2}\sin\left(t\right) & x_{1} & x_{1}x_{2}
\end{array}\right]\theta+u\left(t\right)
\]
where $x:[0,\infty)\rightarrow\mathbb{R}^{2}$, $u:[0,\infty)\rightarrow\mathbb{R}^{2}$,
the unknown parameters were selected as
\[
\theta=\left[\begin{array}{cccc}
5 & 10 & 15 & 20\end{array}\right]^{T},
\]
and the desired trajectory was selected as
\[
x_{d}\left(t\right)=10\left(1-e^{-0.1t}\right)\left[\begin{array}{c}
\sin\left(2t\right)\\
0.4\cos\left(3t\right)
\end{array}\right].
\]

For each of the 200 trials within the Monte Carlo simulation, the
feedback and adaptation gains were selected as $K=K_{s}I_{2}$ and
$\Gamma=\Gamma_{s}I_{4}$, where $K_{s}\in\mathbb{R}$ was sampled
from a uniform distribution on $\left(0.1,\:15\right)$ and $\Gamma_{s}\in\mathbb{R}$
was sampled from a uniform distribution on $\left(0.3,\:3\right)$.
Also, the concurrent learning gain, $k_{CL}$, and the integration
window, $\Delta t$, were sampled from uniform distributions with
support on $\left(0.002,\:0.2\right)$ and $\left(0.01,\:1\right)$,
respectively. After gain sampling, a simulation using each, the traditional
state derivative based, and the integral based, CL update law was
performed, with a step size of 0.0004 seconds and additive white Gaussian
noise on the measured state with standard deviation of 0.3. For each
integral CL simulation, a buffer, with size based on $\Delta t$ and
the step size, was used to store the values of $x$, $Y$, and $u$
during the time interval $\left[t-\Delta t,\:t\right]$ and to calculate
$x\left(t\right)$, $x\left(t-\Delta t\right)$, $\mathcal{Y}\left(t\right)$
and $\mathcal{U}\left(t\right)$. Similarly, for the state derivative
CL simulation, a buffer of the same size was used as the input to
a moving average filter before calculating the state derivative via
central finite difference. The size of the history stack and the simulation
time span were kept constant across all trials at $N=20$ and 100
seconds, respectively.

Since the moving average filter window used in the state derivative
CL simulations provides an extra degree of freedom, the optimal filter
window size was determined\textit{ a priori }for a fair comparison.
The optimal filtering window was calculated by adding Gaussian noise,
with the same standard deviation as in the simulation, to the desired
trajectory, and minimizing the root mean square error between the
estimated and true $\dot{x}_{d}$. This process yielded an optimal
filtering window of 0.5 seconds; however, the filtering window was
truncated to $\Delta t$ on trials where the sampled $\Delta t$ was
less than 0.5 seconds, i.e., $filter\:window=\min\left\{ 0.5,\:\Delta t\right\} $.

\begin{figure}
\begin{centering}
\includegraphics[width=1\columnwidth]{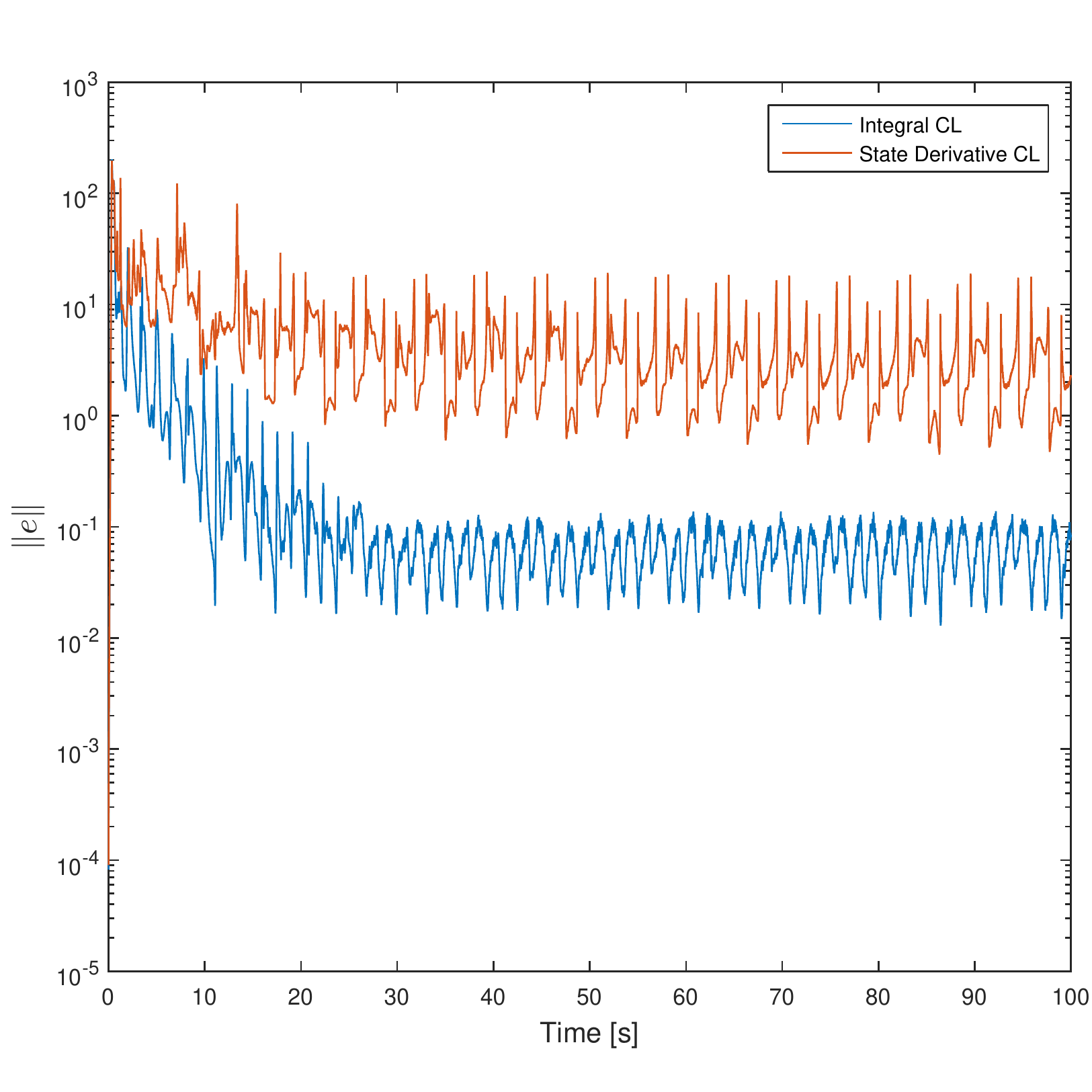}
\par\end{centering}

\caption{\label{fig:error}Mean state trajectory tracking errors across all
trials.}
\end{figure}

\begin{figure}
\begin{centering}
\includegraphics[width=1\columnwidth]{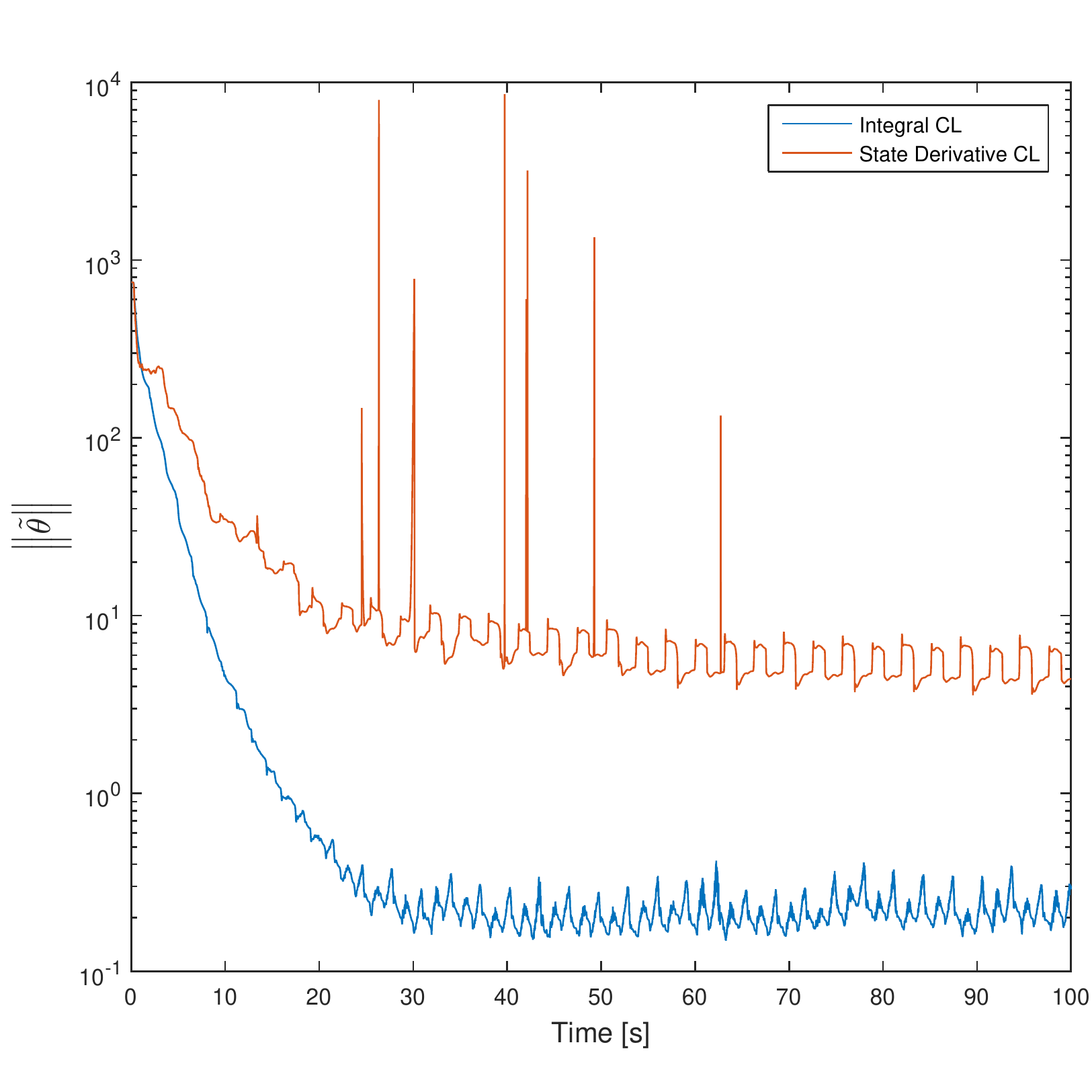}
\par\end{centering}

\caption{\label{fig:thetaError}Mean parameter estimation errors across all
trials.}
\end{figure}

The mean tracking error trajectory and parameter estimation error
trajectory across all trials are depicted in Figs. \ref{fig:error}
and \ref{fig:thetaError}. To compare the overall performance of both
methods, the RMS tracking error and the RMS parameter estimation error
during the time interval $t\in\left[60,\:100\right]$ (i.e., after
reaching steady state) were calculated for each trial, and then the
average RMS errors across all trials was determined. The final results
of the Monte Carlo simulation are shown in Table \ref{tab:simResults},
illustrating the improved performance of integral CL versus state
derivative CL.

\begin{table}
\caption{\label{tab:simResults}Average steady state RMS tracking and RMS parameter
estimation errors across all simulations.}

\centering{}%
\begin{tabular}{|l|c|c|c|c|c|c|}
\cline{2-7} 
\multicolumn{1}{l|}{} &
$e_{1}$ &
$e_{2}$ &
$\theta_{1}$ &
$\theta_{2}$ &
$\theta_{3}$ &
$\theta_{4}$\tabularnewline
\hline 
Integral &
0.1078 &
0.2117 &
.0507 &
0.3100 &
0.1867 &
0.1121\tabularnewline
\hline 
Derivative &
0.2497 &
0.6717 &
0.1802 &
1.3376 &
0.3753 &
0.2382\tabularnewline
\hline 
\end{tabular}
\end{table}

\section{Conclusion\label{sec:Conclusion}}

A modified concurrent learning adaptive update law was developed,
resulting in guarantees on the convergence of the parameter estimation
errors without requiring persistent excitation or the estimation of
state derivatives. The development in this paper represents a significant
improvement in online system identification. Whereas PE is required
in the majority of adaptive methods for parameter estimation convergence
(usually ensured through the use of a probing signal that is not considered
in the Lyapunov analysis), the technique described in this paper does
not require PE. Furthermore, the formulation of concurrent learning
in this paper circumvents the need to estimate the unmeasureable state
derivatives, therefore avoiding the design and tuning of a state derivative
estimator. This formulation is more robust to noise, i.e., has better
tracking and estimation performance, compared to other concurrent
learning designs, as demonstrated by the included Monte Carlo simulation.

\bibliographystyle{IEEEtran}
\bibliography{ncr,master,encr}

\end{document}